\documentstyle[11pt]{article}
\normalbaselines
\newcounter{eqnletter}[equation]
\catcode`\@=11
\@addtoreset{equation}{section}   
\catcode`\@=12

\begin{document}

\begin{center}

{\LARGE\bf Bethe Ansatz in Quantum Mechanics.}\\[0.7cm]
{\Large\bf 1. The Inverse Method of Separation of Variables}
\footnote{This work was partially supported by grant 
436 POL 113/77/0(S) of DFG}

\vskip 1cm

{\large {\bf Dieter Mayer} }

\vskip 0.1 cm

Institute of Theoretical Physics, TU Clausthal,\\
Arnold Sommerfeld Str. 6, 38678 Clausthal, Germany\footnote{E-mail
address: dieter.mayer@tu-clausthal.de}\\

\vskip 0.1cm

{\large and}\\

\vskip 0.1cm

{\large {\bf Alexander Ushveridze} }

\vskip 0.1 cm

Department of Theoretical Physics, University of Lodz,\\
Pomorska 149/153, 90-236 Lodz, Poland\footnote{E-mail
address: alexush@mvii.uni.lodz.pl and alexush@krysia.uni.lodz.pl} \\

\end{center}
\vspace{1 cm}
\begin{abstract}
In this paper we formulate a general method for building completely
integrable quantum systems. The method is based on the use
of the so-called multi-parameter spectral equations, i.e.
equations with several spectral parameters. 
We show that any such equation, after eliminating some spectral
parameters by means of the so-called inverse procedure of
separation of variables can be reduced to a certain completely 
integrable model. Starting with exactly or quasi-exactly
solvable multi-parameter spectral equations we, respectively,
obtain exactly or quasi-exactly solvable integrable models. 
\end{abstract}

\newpage

\section{Introduction}

There is a deep relationship between completely
integrable multi-dimensional quantum systems and
one-dimensional multi-parameter spectral equations. One 
way to explain this relationship is to start
with completely integrable models arising in 
the famous $r$-matrix approach (see e.g. \cite{KoBoIz93,Fa96,Ji90} and
references therein). As it was
demonstrated by Sklyanin, many of these models allow a complete separation of
variables in some generalized coordinate systems which,
generally, cannot be built by means of purely coordinate transformations
(see e.g. \cite{Sk95} and references therein). The result of the
separation of variables is, as a rule, a system of one-dimensional
differential (or pseudo-differential) equations with several 
spectral parameters. One of these parameters is the energy of the 
initial system, while the others play the role of so-called 
{\it separation constants}. The spectral problem for the initial model 
reduces then to the problem of finding the admissible values for these 
parameters for which the separated equations have solutions of a
special form. This form is dictated by the structure of the
Hilbert space of the initial model and can be considered as an
{\it ansatz} for the resulting equations. The general structure of this 
ansatz is closely related to the famous {\it Bethe Ansatz}
which can be used for constructing solutions of the initial
completely integrable system. It also turns out that the numerical equations 
determining the spectra of the multi-parameter spectral equations 
exactly coincide with the so-called Bethe Ansatz equations determining the
spectrum of the initial model.

These facts suggest to formulate an
independent approach for building completely integrable and
exactly solvable quantum systems.
The idea is very simple: One should start with a certain
$N$-parameter spectral equation having an exactly constructable set of 
solutions and interpret $N$ copies of it (rewritten in $N$
different variables) as the result of separation of
variables in a certain $N$-dimensional quantum model.
Identifying one of the $N$ spectral parameters with the
energy of this $N$-dimensional model and considering the remaining 
$N-1$ parameters as separation constants, one can easily reconstruct the
form of its hamiltonian. Since the total number of spectral
parameters is $N$, there are $N$ different ways for
identifying a spectral parameter with the energy. 
This leads us to $N$ independent hamiltonians whose spectra
are completely determined by the spectrum of the initial
$N$-parameter spectral equation. It can also be
shown that the resulting multi-dimensional
hamiltonians commute with each other and thus form a
certain $N$-dimensional completely integrable (and exactly
solvable) system. It is worth stressing that this (inverse) 
method is far from being new: It appeared approximately at the same time as
Sklyanins (direct) method (see ref. \cite{Sk87}) and was exposed 
in refs. \cite{Us88,Us89,Us92,Us94}
where it was called the {\it inverse method of separation of
variables.} 

With this paper we start a series of publications devoted
to a detailed exposition of this inverse method of separation
of variables. We intend to consider this method as a possible alternative
(or rather complement) to the existing $r$-matrix approach.
In our opinion, this method has several essential advantages 
in comparison with the $r$-matrix approach: First of all, the problem
of constructing exactly solvable multi-parameter spectral
equations is rather simple and leads to a very rich class
of such equations. Applying to these equations the inverse method
we can construct the corresponding completely
integrable systems. It is easily seen that the class of
systems obtained in this way is considerably richer
than that obtained in the fraework of the
standard $r$-matrix approach. The second advantage lies in the
fact that we can consider not only exactly but also the
so-called quasi-exactly solvable multi-parameter spectral equations.
These equations are distinguished by the fact that they can
be solved by purely algebraic methods only for some limited
parts of the spectrum but not for the whole spectrum. The
completely integrable systems obtained this way also
turn out to be {\it quasi-exactly solvable} (about
quasi-exactly solvable models and methods for their
construction see e.g. the review articles 
\cite{Us89,Sh89,MPRST90,Us92,Sh94,GKO94} and the
book \cite{Us94} ). As far as we
know, at present there are no methods for building
such systems within the framework of the $r$-matrix approach.
The third advantage finally is that the method for building exactly
solvable multi-parameter spectral equations gives
automatically the form of their solutions. This means that
the problem of solving the resulting multi-dimensional
completely integrable systems does not appear at all. We
know these solutions from the very beginning! This is not the
case for the $r$-matrix approach where the construction of
explicit solutions for completely integrable models requires
considerable efforts and is not formalised yet.

Of course, there are also some difficulties with the
inverse method of separation of variables, the most obvious of which can
be formulated as follows: The completely integrable models
obtained by the method are formulated 
in those coordinate systems in which they
are separable. Often it is desirable to rewrite them in
a form which is typical for the $r$-matrix approach,
i.e. as elements of the universal enveloping algebras
of some Lie algebras (we think here about various spin chains and
their generalizations). Unfortunately, this problem is not
so easy, and, at present, it is solved only for a few
cases of models.

\section{Multi-parameter spectral equations}

\subsection{General definitions}

The {\it multi-parameter spectral equations} play a
fundamental role in our approach. We start this section
giving their precise mathematical definition.

\medskip
Let $V_i,\ i=1,\ldots,N$ be complex linear vector 
spaces of dimension $d_i$ which we call {\it carrier spaces}. 
For linear operators 
\begin{eqnarray}
A_i \in\mbox{End}\ [V_i], \qquad i=1,\ldots, N
\label{mps.1}
\end{eqnarray}
and
\begin{eqnarray}
B_i^j \in\mbox{End}\ [V_i], \qquad j=1,\ldots,N, \quad i=1,\ldots, N
\label{mps.2}
\end{eqnarray}
consider the problem of finding all possible 
sets of complex numbers $\lambda_1,\ldots,\lambda_N$ for which the relations
\begin{eqnarray}
\left[A_i-\sum_{j=1}^N B_i^j \lambda_j\right]\phi_i=0, 
\qquad i=1,\ldots,N
\label{mps.3}
\end{eqnarray}
hold  for some $V_i\ni \phi_i,\ \phi_i \neq 0$.
We consider these relations as a system of simultaneous 
equations for the numbers $\lambda_1,\ldots,\lambda_N$ and vectors 
$\phi_1,\ldots,\phi_N$.

\medskip
{\bf Definition 1.1.} We call the system (\ref{mps.3}) a
system of multi-parameter spectral equations.
The numbers $\lambda_1,\ldots,\lambda_N$ will be called 
{\it spectral parameters}.

\medskip
It is convenient to regard the parameters $\lambda_1,\ldots,\lambda_N$
as coordinates of a point in a certain affine space
$\Lambda$ of dimension $N$. 

\medskip
{\bf Definition 1.2.} 
We call $\Lambda$ the {\it spectral space} and those of its
points for which the system (\ref{mps.3}) becomes solvable
we call the {\it spectral points}. The set of all spectral
points will be called the {\it spectrum} of system
(\ref{mps.3}) and is denoted by $\Sigma$. 

\medskip
Obviously, the spectrum $\Sigma$ 
can be found from the system of secular equations
\begin{eqnarray}
{\det}_i\left[A_i-\sum_{j=1}^N B_i^j\lambda_j\right]=0,\qquad
i=1,\ldots, N
\label{mps.3a}
\end{eqnarray}
in which ${\det}_i$ denotes the deteminant of operators
acting in the space $V_i$. Since (\ref{mps.3a}) is a
system of $N$ equations for $N$ unknowns, the spectrum of
system (\ref{mps.3}) is generally discrete, except for some
special degenerate cases when it may be continuous or empty. 

\medskip
{\bf Definition 1.3.} We call a system of multi-parameter spectral
equations {\it finite - dimensional} 
if the corresponding
carrier spaces are all finite - dimensional. Otherwise it
is {\it infinite - dimensional}.

\medskip
In the general case, the number of solutions of a
finite-dimensional system of multi-parameter
spectral equations is 
\begin{eqnarray}
N_{sol}=\prod_{i=1}^N \dim V_i
\label{mps.3aaa}
\end{eqnarray}
Indeed, if all the carrier spaces $V_i$ are finite-dimensional,
the secular equations
(\ref{mps.3a}) for the numbers $\lambda_1,\ldots,\lambda_N$ 
become algebraic equations  of orders $\dim V_i,\ i=1,\ldots, N$.
The number of solutions for a system of such equations is, obviously,
given by formula (\ref{mps.3aaa}).
If the dimensions of carrier spaces tend to infinity, the number of
solutions of system (\ref{mps.3a}) increases. In
the limit $\dim V_i=\infty$ the spectrum of the multi-parameter
spectral equations (\ref{mps.3}) becomes infinite.
This leads us to the following simple result:

\medskip
{\bf Proposition 1.1.} Systems of finite-dimensional multi-parameter
spectral equations have finite spectrum. The spectra of 
infinite-dimensional systems of multi-parameter spectral equations are
in general infinite.

\subsection{Two trivial cases}

In particular cases system (\ref{mps.3}) reduces to 
well known equations of linear algebra. Consider the
following two simple examples.

\medskip
{\bf Example 1.1.} Let $N>1$ and $\dim V_i=1,\
i=1,\ldots,N$. Then the operators $A_i$ and $B_i^j$ become
numbers. The role of the vectors $\phi_i$ is played by
non-zero numbers which can be chosen arbitrarily because of
the linearity of (\ref{mps.3}). 
In this case the spectral parameters  $\lambda_i$ satisfy a system of 
inhomogeneous linear equations of the form
\begin{eqnarray}
A_i-\sum_{j=1}^N B_i^j \lambda_j=0, 
\qquad i=1,\ldots,N.
\label{mps.3b}
\end{eqnarray}
If the matrix $B_i^j$ is non-degenerate, then
(\ref{mps.3b}) has an unique solution
\begin{eqnarray}
\lambda_j=\sum_{i=1}^N (B^{-1})_j^i A_i, 
\qquad j=1,\ldots,N
\label{mps.3c}
\end{eqnarray}
and the spectrum of system (\ref{mps.3}) consists of a
single spectral point. 

\medskip
{\bf Example 1.2.} Let $N=1$ and $\dim V_1> 1$. We denote
$V_1=V$, $A_1=A$, $B_1^1=B$ and also $\phi_1=\phi$ and $\lambda_1=\lambda$. 
Then system (\ref{mps.3}) reduces to a single one-parameter
spectral equation of the form
\begin{eqnarray}
\left[A-B \lambda\right]\phi=0, 
\qquad V\ni \phi \neq 0.
\label{mps.3d}
\end{eqnarray}
If the operator $B$ is non-degenerate, then 
\begin{eqnarray}
\lambda\in \mbox{Spec}\left[B^{-1}A\right].
\label{mps.3e}
\end{eqnarray}
In this case the properties of the spectrum of system (\ref{mps.3})
can be analysed simply by using the standard
spectral theory of linear operators in vector spaces.

\medskip
The two examples demonstrate that
multi-parameter spectral equations of the form
(\ref{mps.3}) can be
regarded as generalizations of both the inhomogeneous
(non-spectral) and homogeneous (spectral) linear equations.

\subsection{More complicated examples}

It is perhaps instructive to consider two
simple examples in which the features of both the
inhomogeneous and homogeneous equations are
simultaneously present. Let us first consider an example of
finite-dimensional multi-parameter spectral equations.

\medskip
{\bf Example 1.3.} Let $N=2$ and $\dim V_1=\dim V_2=2$. In
this case all the operators $A_1,\ B_1^1,\ B_1^2$ and 
$A_2,\ B_2^1,\ B_2^2$ can be represented by $2\times 2$ matrices
acting on two-dimensional vectors $\phi_1$ and $\phi_2$.
The two spectral parameters $\lambda_1$ and $\lambda_2$ can
be found from the system of two secular equations
\begin{eqnarray}
\det\left(A_1-B_1^1\lambda_1-B_1^2\lambda_2\right)=0,\qquad
\det\left(A_2-B_2^1\lambda_1-B_2^2\lambda_2\right)=0,
\label{mps.3gg}
\end{eqnarray}
which both are algebraic equations of order two. Hence,
system (\ref{mps.3gg}) has four solutions, and therefore
the spectrum of the corresponding multi-parameter
spectral equations consists of four spectral points.
The construction of the vectors $\phi_1$ and $\phi_2$
associated with these spectral points can be performed in a
standard way.

The next example concerns an infinite-dimensional system of multi-parameter 
spectral equations.

\medskip
{\bf Example 1.4.} Let $N=2$. Let $V_1$ and $V_2$ be the
spaces of analytic functions of a real variable $x$
vanishing at the ends of the 
intervals $[a_1,b_1]$ and $[a_2,b_2]$, respectively.
Then, obviously, $\dim V_1=\dim V_2=\infty$. Let us take
\begin{eqnarray}
A_1=A_2=\frac{\partial^2}{\partial x^2}, \qquad B_i^j=\omega_i^j(x),
\quad i,j=1,2
\label{mps.3f}
\end{eqnarray}
where $\omega_i^j(x)$ are certain analytic functions of $x$
on the corresponding intervals.
In this case, the system (\ref{mps.3}) takes the form
\begin{eqnarray}
\begin{array}{lll}
\left[\frac{\partial^2}{\partial x^2}-\omega_1^1(x)\lambda_1-
\omega_1^2(x)\lambda_2
\right]\phi_1(x)=0, \quad & \phi_1(x)\not\equiv 0,\quad &
\phi_1(a_1)=\phi_1(b_1)=0,\\[0.2cm]
\left[\frac{\partial^2}{\partial x^2}-\omega_2^1(x)\lambda_1-
\omega_2^2(x)\lambda_2
\right]\phi_2(x)=0, \quad & \phi_2(x)\not\equiv 0,\quad &
\phi_2(a_2)=\phi_2(b_2)=0.
\end{array}
\label{mps.3g}
\end{eqnarray}
The spectrum of this system can be found from the system of
numerical equations
\begin{eqnarray}
\begin{array}{cc}
\Phi_1(\lambda_1,\lambda_2, c_1, a_1)=0,\quad &
\Phi_1(\lambda_1,\lambda_2, c_1, b_1)=0,\\
\Phi_2(\lambda_1,\lambda_2, c_2, a_2)=0,\quad &
\Phi_2(\lambda_1,\lambda_2, c_2, b_2)=0,
\end{array}
\label{mps.3h}
\end{eqnarray}
in which $\Phi_1(\lambda_1,\lambda_2,c_1,x)$ and 
$\Phi_2(\lambda_1,\lambda_2,c_2,x)$
denote the general (normalized) solutions of equations (\ref{mps.3g}).
The system (\ref{mps.3h}) is transcendental and generally
has an infinite and discrete set of solutions. Thus the
spectrum of system (\ref{mps.3g}) is infinite and discrete.

\subsection{Equivalence transformations}

Let $S_i\in\mbox{End}\ [V_i]$ and $R_i\in\mbox{End}\ [V_i]$
be arbitrary invertible operators, $M_i^k$ be the
elements of an arbitrary
non-degenerate $N\times N$ complex matrix and $\mu_i$ be 
arbitrary complex numbers. It is not difficult to see that 
the transformations of operators $A_i$ and $B_i^j$ defined as
\begin{eqnarray}
\tilde A_i=S_i \left[A_i-\sum_{k=1}^N B_i^k\mu_k\right]R^{-1}_i 
\in \mbox{End}\ [V_i],\qquad i=1,\ldots,N,
\label{mps.3i}
\end{eqnarray}
and
\begin{eqnarray}
\tilde B_i^j = S_i \left[\sum_{k=1}^N B_i^k (M^{-1})_k^j\right]
R^{-1}_i\in\mbox{End}\ [V_i], \qquad j=1,\ldots,N, \quad i=1,\ldots, N,
\label{mps.3j}
\end{eqnarray}
and the simultaneous transformations of the spectral parameters
$\lambda_i$
\begin{eqnarray}
\tilde\lambda_j=\sum_{k=1}^N M_j^k(\lambda_k-\mu_k),  
\label{mps.3k}
\end{eqnarray}
and vectors $\phi_i$
\begin{eqnarray}
\tilde\phi_i=R_i\phi_i 
\label{mps.3kk}
\end{eqnarray}
leave the form of equations (\ref{mps.3}) unchanged:
\begin{eqnarray}
\left[\tilde A_i-\sum_{j=1}^N \tilde B_i^j \tilde\lambda_j\right]
\tilde\phi_i=0, 
\qquad V_i\ni \tilde\phi_i \neq 0,\qquad i=1,\ldots,N.
\label{mps.3l}
\end{eqnarray}

\medskip
{\bf Definition 1.4.} The transformations (\ref{mps.3i}) -- (\ref{mps.3kk}) 
will be called {\it equivalence transformations} of
system (\ref{mps.3}). The systems (\ref{mps.3}) and
(\ref{mps.3l}) related by such transformations will be
called {\it equivalent}. We shall distinguish three particular
cases of such transformations: 

1) transformations of first kind:
$S_i\neq I_i,\ R_i=I_i,\  
M_i^k=\delta_i^k,\ \mu_i=0,\ i=1,\dots, N$, 

2) transformations of second kind: 
$S_i= I_i,\ R_i\neq I_i,\  
M_i^k=\delta_i^k,\ \mu_i=0,\ i=1,\dots, N$, 

3) transformations of third kind: 
$S_i= I_i,\ R_i=I_i,\  M_i^k\neq \delta_i^k,
\ \mu_i\neq 0,\ i=1,\dots, N$. \\
where $I_i$ denotes the unit operator acting in
the carrier space $V_i$. 

\medskip
It can be seen that the transformations of the first kind 
do not change the solutions of system (\ref{mps.3}), while
the transformations of second and third kind transform the
sets of spectral parameters $\lambda_j$
and vectors $\phi_i$, respectively. Note also that the
transformations of second kind are ordinary linear
transformations in the carrier spaces $V_i$, while the
transformations of third kind are affine
transformations in the spectral space $\Lambda$.

\subsection{Degeneracy of the spectrum $\Sigma$}

As noted above, the spectrum of multi-parameter spectral
equations can be regarded as a set of points in a
$N$-dimensional affine space $\Lambda$. The form of this set
strongly depends on the concrete form of the system under consideration.
In particular, it may happen that the spectrum of a certain system
can be imbedded into another affine space $\Lambda'$ of
smaller dimension $N'< N$. In this case we shall say that
this system of multi-parameter spectral equations has a
{\it completely degenerate} spectrum. We can also imagine the 
situation where not the whole spectrum of a system but only
some part of it can be imbedded into a space $\Lambda'$
of smaller dimension. Such a 
spectrum will be called {\it partially degenerate}.
Below we give a little more precise definition of degeneracy.

\medskip
{\bf Definition 1.5.} Let $\Sigma' \subset \Sigma$ denote
some subset of the spectrum of some multi-parameter spectral equations. 
We call this subset {\it degenerate} if there exists a rectangular
$K\times N$ matrix $M_i^j,\ i=N-K+1,\ldots, N, \ j=1,\ldots, N$
with $K<N$ linearly independent rows and a $K$-dimensional
vector $\mu_i,\ i=N-K+1,\ldots, N$, such that
\begin{eqnarray}
\sum_{j=1}^N M_i^j(\lambda_j-\mu_i)=0,\qquad i=N-K+1,\ldots, N
\label{mps.4aa}
\end{eqnarray}
for any $(\lambda_1,\ldots,\lambda_N)\in \Sigma'$. 
The number $K$ (which is equal to the rank of the matrix $M_i^j$)
will be called the {\it degree} of the degeneracy. We call
a degenerate subset $\Sigma'$ {\it non-extendable} if there
are no other elements of the spectrum $\Sigma$ which  
satisfy relations (\ref{mps.4aa}) with the same $M_i^j$ and $\mu_i$.

\medskip
It is not difficult to see that formula (\ref{mps.4aa})
describes part of an equivalence transformation of
third kind described in the previous subsection (see
formula (\ref{mps.3k})). This enables one to see that the
degeneracy of a set $\Sigma'\subset\Sigma$ implies 
the existence of equivalence 
transformations of third kind which bring the values of $K$ spectral 
parameters on $\Sigma'\subset \Sigma$ to zero. 

\medskip
{\bf Definition 1.6.} We call the spectrum $\Sigma$ of a system 
of multi-parameter spectral equations {\it completely degenerate} 
if $\Sigma$ itself is degenerate and {\it partially degenerate} if 
$\Sigma$ contains a finite, degenerate and non-extendable 
subset $\Sigma'$.

\subsection{Hermitian multi-parameter spectral equations}

Assume now that the carrier spaces $V_i$ are Hilbert spaces
endowed with scalar products $(\ ,\ )_i$.

\medskip
{\bf Definition 1.8.} We call a system of multi-parameter
spectral equations {\it hermitian} if, for all $i=1,\ldots,
N$, the operators $A_i$ and $B_i^j$ are hermitian 
in the corresponding spaces $V_i$. 

\medskip
Among many nice properties of 
systems of hermitian multi-parameter spectral equations
there are two which play an especially 
important role in many applications. In order to derive them
assume system (\ref{mps.3}) to be hermitian. Then, taking
the scalar product of the $i$th equation of this system with $\phi_i$, 
we obtain a system of ordinary linear inhomogeneous equations
\begin{eqnarray}
\bar A_i-\sum_{j=1}^N \bar B_i^j\lambda_i=0,\qquad i=1,\ldots, N
\label{mpk.1}
\end{eqnarray}
with $\bar A_i=(\phi_i,A_i\phi_i)_i$ and 
$\bar B_i^j=(\phi_i,B_i^j\phi_i)_i$. Hermiticity of the
operators $A_i$ and $B_i^j$ implies the numbers
$\bar A_i$ and $\bar B_i^j$ to be real. But this means that the
solutions $\lambda_j$ of system (\ref{mpk.1}) also must be real.
Thus we arrive at the following result:

\medskip
{\bf Proposition 1.3.} The spectra of the systems of hermitian
multi-parameter spectral equations are real.

\medskip
Let $\phi_i^{(n)}$ and $\phi_i^{(m)}$, $i=1,\ldots,N$ be
two different solutions of system (\ref{mps.3}) with the corresponding
sets of spectral parameters $\lambda_j^{(n)}$ and $\lambda_j^{(m)}$,
$j=1,\ldots,N$. Then one has
\begin{eqnarray}
\left[A_i-\sum_{j=1}^N B_i^j \lambda_j^{(n)}\right]\phi_i^{(n)}=0, 
\qquad \phi_i^{(n)}\in V_i,\qquad i=1,\ldots,N
\label{mpk.2}
\end{eqnarray}
and
\begin{eqnarray}
\left[A_i-\sum_{j=1}^N B_i^j \lambda_j^{(m)}\right]\phi_i^{(m)}=0, 
\qquad \phi_i^{(m)}\in V_i,\qquad i=1,\ldots,N
\label{mpk.3}
\end{eqnarray}
Taking the scalar product of (\ref{mpk.2}) with $\phi_i^{(m)}$ and
(\ref{mpk.3}) with $\phi_i^{(n)}$, we obtain
\begin{eqnarray}
\left(\phi_i^{(m)},A_i\phi_i^{(n)}\right)_i = 
\sum_{j=1}^N \left(\phi_i^{(m)},B_i^j\phi_i^{(n)}\right)_i
\lambda_j^{(n)}, 
\qquad i=1,\ldots,N
\label{mpk.4}
\end{eqnarray}
and
\begin{eqnarray}
\left(\phi_i^{(n)},A_i\phi_i^{(m)}\right)_i = 
\sum_{j=1}^N \left(\phi_i^{(n)},B_i^j\phi_i^{(m)}\right)_i
\lambda_j^{(m)}, 
\qquad i=1,\ldots,N.
\label{mpk.5}
\end{eqnarray}
Due to hermiticity of the operators $A_i$ and $B_i^j$ in
the spaces $V_i$, we get by subtracting (\ref{mpk.4}) from (\ref{mpk.5})
\begin{eqnarray}
\sum_{j=1}^N \left(\phi_i^{(m)},B_i^j\phi_i^{(n)}\right)_i
(\lambda_j^{(m)}-\lambda_j^{(n)})=0, 
\qquad i=1,\ldots,N.
\label{mpk.6}
\end{eqnarray}
Let us now assume that the spectral parameters are not degenerate.
Then
equations (\ref{mpk.6}) can be satisfied if and only if 
\begin{eqnarray}
\det \left|\left|\left(\phi_i^{(m)},B_i^j\phi_i^{(n)}\right)_i
\right|\right|_{i,j=1}^N=0. 
\label{mpk.7}
\end{eqnarray}

\medskip
{\bf Definition 1.9.}
We shall say that two solutions $\phi_i^{(n)}$ and
$\phi_i^{(m)}$, $i=1,\ldots, N$ of system (\ref{mps.3}) are
{\it orthogonal} (in the generalized sense) if they satisfy
condition (\ref{mpk.7}) which we call the {\it
generalized orthogonality condition}.

\medskip
{\bf Proposition 1.4.} The solutions of a system of hermitian
multi-parameter spectral equations corresponding to
different spectral points are orthogonal in
the generalized sense.

\medskip
It is easily seen that Propositions 1.3 and 1.4 are  natural
generalizations of well known results in linear algebra
about reality of spectra of hermitian operators in Hilbert
spaces and orthogonality of eigensolutions corresponding to
different eigenvalues.

\section{The inverse method of separation of variables}

\subsection{Preliminary steps}

We will show that any multi-parameter spectral
equations can be reduced to a system of ordinary spectral
equations with a single spectral parameter.
This can be done by means of the so-called {\it inverse
method of separation of variables} (see e.g. refs. \cite{}). 
To explain this consider the
{\it composite carrier space}
\begin{eqnarray}
{\cal V}=V_1\otimes\ldots\otimes V_N
\label{mps.4}
\end{eqnarray}
and introduce the operators
\begin{eqnarray}
{\cal A}_i=I_1\otimes\ldots\otimes A_i \otimes\ldots\otimes I_N 
\in \mbox{End}\ [{\cal V}], \qquad i=1,\ldots,N
\label{mps.5}
\end{eqnarray}
respectively
\begin{eqnarray}
{\cal B}_i^j=I_1\otimes\ldots\otimes B_i^j \otimes\ldots\otimes I_N 
\in \mbox{End}\ [{\cal V}], \qquad j=1,\ldots, N,\quad i=1,\ldots,N
\label{mps.5a}
\end{eqnarray}
acting in ${\cal V}$. Here as before
$I_i\in \mbox{End}\ [V_i]$ denotes the unit operator in $V_i$.
Taking  
\begin{eqnarray}
\varphi=\phi_1\otimes\ldots\otimes\phi_N \in {\cal V},
\label{mps.6}
\end{eqnarray}
we can rewrite system (\ref{mps.3}) in the form
\begin{eqnarray}
\left[{\cal A}_i-\sum_{j=1}^N {\cal B}_i^j \lambda_j\right]
\varphi=0, \quad {\cal V}\ni \varphi \neq 0, \qquad i=1,\ldots, N.
\label{mps.7}
\end{eqnarray}
This means that any solution of the initial system (\ref{mps.3})
generates a solution of system (\ref{mps.7}).

Now consider the $N\times N$ matrix ${\cal B}_i^j$ in equation 
(\ref{mps.7}). From its definition (\ref{mps.5a}) it follows
that the entries of 
this matrix belonging to different rows labeled by the index $i$
commute with each other. Hence, the matrix determinant of this 
operator-valued matrix 
\begin{eqnarray}
\rho=\sum_{j_1,\ldots, j_N=1}^N \varepsilon_{j_1,\ldots,j_N}
{\cal B}^{j_1}_1\cdot\ldots\cdot {\cal B}^{j_N}_N 
\label{mps.8}
\end{eqnarray}
is well defined and does not depend on the order of the factors in 
formula (\ref{mps.8}). For this determinant, which defines obviously 
an operator, one can write
\begin{eqnarray}
\delta_k^j\ \rho =
\sum_{i=1}^N\Delta{\cal B}_k^i {\cal B}_i^j
\label{mps.9}
\end{eqnarray}
with
\begin{eqnarray}
\Delta{\cal B}_j^i=
\sum_{j_1,\ldots,j_{i-1}, j_{i+1},\ldots, j_N=1}^N
\varepsilon_{j_1,\ldots,j_{i-1},j,j_{i+1},\ldots,j_N}
{\cal B}^{j_1}_1\ldots {\cal B}^{j_{i-1}}_{i-1} 
{\cal B}^{j_{i+1}}_{i+1},\ldots {\cal B}^{j_N}_N
\label{mps.10}
\end{eqnarray}
the cofactors of the elements ${\cal B}_i^j$.
Acting by this operator matrix (\ref{mps.10}) on both sides
of (\ref{mps.7}), and introducing the notations 
\begin{eqnarray}
{\cal C}_j=
\sum_{i=1}^N\Delta{\cal B}_j^i {\cal A}_i, \qquad i=1,\ldots,N
\label{mps.12}
\end{eqnarray}
we obtain
\begin{eqnarray}
{\cal C}_j\varphi=\lambda_j\rho\varphi, \qquad \varphi\in {\cal
V},\qquad j=1,\ldots,N
\label{mps.11}
\end{eqnarray}
Hence, we arrive at the following result.

\medskip
{\bf Proposition 2.1.} Any solution of the system of multi-parameter 
spectral equations (\ref{mps.3}) generates a certain solution of the 
system of one-parameter spectral equations (\ref{mps.11}).

\medskip
It is not difficult to see that the transition from (\ref{mps.3})
to (\ref{mps.11}) essentially solves the inverse problem of
separation of variables in a very abstract formulation. The
system of ``one-dimensional'' $N$-parameter spectral equations (\ref{mps.3})
is interpreted as a system of equations appearing after the
separation of ``variables'' in a certain
``$N$-dimensional'' one-parameter spectral equation. 
One of the $N$ spectral parameters entering system (\ref{mps.3}),
say parameter $\lambda_j$, is identified
with a single spectral parameter of this equation.
The $N-1$ remaining spectral parameters $\lambda_k,\ k\neq j$ 
take the role of separation constants. The
reconstruction of the form of the separable
equation is performed by eliminating the separation
constants from system (\ref{mps.3}). Since the index $j$ may
take $N$ different values from $1$ to $N$, we, in fact, can
associate with system (\ref{mps.3}) $N$ different separable equations
which are listed in formula (\ref{mps.11}). 

\medskip
{\bf Definition 2.1.} The equations (\ref{mps.11}) will be
called the {\it universal separable equations}. The
operator $\rho$ in this equation will be called the {\it
weight operator}. 

\medskip
The universality of equations (\ref{mps.11}) is reflected by the
fact that they can be derived for {\it arbitrary} systems of
multi-parameter spectral equations 
(\ref{mps.3}) irrespective of the concrete form and properties of
the operators $A_i$ and $B_i^j$ forming this system.

\subsection{Transformation properties of the universal
separable equations}

In this subsection we derive the form of the universal separable 
equations (\ref{mps.11}) constructed from the transformed
equations (\ref{mps.3l}). Introducing the operators
\begin{eqnarray}
{\cal S}=S_1\otimes\ldots\otimes S_N 
\in \mbox{End}\ [{\cal V}], \qquad
{\cal R}=R_1\otimes\ldots\otimes R_N 
\in \mbox{End}\ [{\cal V}],
\label{mps.12a}
\end{eqnarray}
and using formulas (\ref{mps.8}) -- (\ref{mps.12}),
it is not difficult to check that the transformed version
of system (\ref{mps.11}) reads
\begin{eqnarray}
\tilde{\cal C}_j\tilde\varphi=\tilde\lambda_j\tilde\rho\tilde\varphi, 
\qquad \tilde\varphi\in {\cal
V},\qquad j=1,\ldots,N
\label{mps.12b}
\end{eqnarray}
where
\begin{eqnarray}
\tilde{\cal C}_j=
{\cal S}\left[\sum_{i,k=1}^N M_j^k\left(\Delta{\cal B}_k^i 
{\cal A}_i-\mu_j I\right)\right]{\cal R}^{-1},
\label{mps.12c}
\end{eqnarray}
\begin{eqnarray}
\tilde\rho={\cal S}\rho{\cal R}^{-1},
\label{mps.12d}
\end{eqnarray}
and 
\begin{eqnarray}
\tilde\lambda_j=\sum_{k=1}^N M_j^k(\lambda_k-\mu_k),  
\label{mps.3ku}
\end{eqnarray}
respectively
\begin{eqnarray}
\tilde\varphi={\cal R}\varphi.
\label{mps.3kku}
\end{eqnarray}
Here we denoted by $I$ the unit operator in the composite
carrier space ${\cal V}$.

\subsection{Transition to completely integrable models}

Up to now we did not assume any special properties for the operators
$A_i$ and $B_i^j$ in
(\ref{mps.1}) and (\ref{mps.2}). In this section we show,
that there exists a wide class of these operators for which 
the corresponding universal separable equations (\ref{mps.11}) 
can be interpreted as
spectral equations for a certain completely integrable
quantum system. We start with the following definition:

\medskip
{\bf Definition 2.2.} We call the system of multi - parameter
spectral equations non - degenerate if its weight operator $\rho$
defined by formula (\ref{mps.8}) is non-singular, i.e. if
$\det \rho\neq 0$.

\medskip
Invertibility of the weight operator implies the existence of operators
\begin{eqnarray}
{\cal D}_i=\rho^{-1}{\cal C}_i\in \mbox{End}\ [{\cal V}],\qquad i=1,\ldots,r
\label{im.1}
\end{eqnarray}
whose eigenvalues in the space ${\cal V}$ coincide with the
values of the parameters $\lambda_i$:
\begin{eqnarray}
{\cal D}_j\varphi=\lambda_j\varphi, \qquad \varphi\in {\cal
V},\qquad j=1,\ldots,N
\label{im.2}
\end{eqnarray}
From (\ref{im.2}) it immediately follows that 
\begin{eqnarray}
[{\cal D}_j, {\cal D}_k]\varphi=0, \qquad \varphi\in {\cal
V},\qquad j,k=1,\ldots,N
\label{im.3}
\end{eqnarray}
for any $\varphi$ satisfying equations (\ref{im.2}). 
This means that the operators ${\cal D}_j$ commute with each other in
the week sence, i.e. on the states $\varphi$ built from the
solutions of the initial system (\ref{mps.3}).
Commutativity in the strong sense does not follow generally
from (\ref{im.3}) if it is not {\it a priori} known
that the set of these states is complete in ${\cal V}$. 

Fortunately, there exists a special case in which strong
commutativity of operators ${\cal D}_j$ can be guaranteed
without any knowledge of the completeness of $\varphi$'s.
This case happens when the
operators $B_i^j,\ j=1,\ldots,N$ 
commute with each other for fixed $i=1,\ldots,N$:
\begin{eqnarray}
[B_i^j, B_i^k]=0, \qquad i,j,k=1,\ldots,N.
\label{im.4}
\end{eqnarray}

\medskip
{\bf Definition 2.3.} We call a system of multi-parameter
spectral equations {\it commutative} if the operators
$B_i^j$ in (\ref{mps.2})
satisfy  relations (\ref{im.4}).

\medskip
In order to demonstrate that commutativity of system
(\ref{mps.3}) implies commutativity of the operators ${\cal
D}_i$ in (\ref{im.1}), 
let us first note that if the
operator $\rho$ is invertible, the operators 
${\cal D}_j$ can be defined as solutions of the system of linear operator
equations
\begin{eqnarray}
{\cal A}_i=\sum_{j=1}^N {\cal B}_i^j{\cal D}_j, \qquad i=1,\ldots,N.
\label{im.5}
\end{eqnarray}
From definitions (\ref{mps.5}) and (\ref{mps.5a}) of the
operators ${\cal A}_i$ and ${\cal B}_i^j$ 
and formula (\ref{im.4}) it follows
that 
\begin{eqnarray}
[{\cal A}_i,{\cal A}_k]=0, \quad \mbox{for any}\quad i, k,
\label{im.6}
\end{eqnarray}
\begin{eqnarray}
[{\cal B}_i^j,{\cal B}_k^l]=0, \quad \mbox{for all}\quad i,k,j,l
\label{im.7}
\end{eqnarray}
\begin{eqnarray}
[{\cal A}_i,{\cal B}_k^j]=0, \quad \mbox{for any}\quad
i\neq k\quad \mbox{and any}\quad j.
\label{im.8}
\end{eqnarray}
Taking into account relation (\ref{im.5}), we get the following chain
of equalities:
\begin{eqnarray}
\sum\limits_{n,m=1}^N{\cal B}_i^n{\cal B}_i^m\big[{\cal D}_n,{\cal D}_m\big]=
\sum\limits_{n,m=1}^N{\cal B}_i^n\big[{\cal D}_n,{\cal B}_i^m\big]{\cal D}_m+
\sum\limits_{n,m=1}^N{\cal B}_i^m\big[{\cal B}_i^n,{\cal D}_m\big]{\cal D}_n+
\nonumber\\
+\sum\limits_{n,m=1}^N\big[{\cal B}_i^n,{\cal B}_i^m\big]{\cal
D}_n {\cal D}_m+
\sum\limits_{n,m=1}^N{\cal B}_i^n{\cal B}_i^m\big[{\cal D}_n,{\cal D}_m\big]=
\sum\limits_{n,m=1}^N\big[{\cal B}_i^n {\cal D}_n,{\cal B}_i^m 
{\cal D}_m\big]=
\nonumber\\
=\big[{\cal A}_i,{\cal A}_i\big]=0.
\label{im.9}
\end{eqnarray}
Here we used the fact that the last two terms in the first 
line of (\ref{im.9}) cancel
and the first term in the second line 
vanishes because of the commutation relation 
(\ref{im.7}). 
Now, using relations (\ref{im.6}) and (\ref{im.8}), we can consider an
analogous chain for $i\neq k$:
\begin{eqnarray}
\sum\limits_{n,m=1}^N{\cal B}_i^n{\cal B}_k^m \big[{\cal D}_n,
{\cal D}_m\big]=
\sum\limits_{n,m=1}^N{\cal B}_i^n{\cal B}_k^m \big[{\cal D}_n,
{\cal D}_m\big]+ 
\sum\limits_{m=1}^N \big[{\cal B}_i^0,{\cal B}_k^m\big]{\cal D}_m=
\nonumber\\
=\sum\limits_{n,m=1}^N{\cal B}_i^n{\cal B}_k^m \big[{\cal D}_n,
{\cal D}_m\big]+
\sum\limits_{n,m=1}^N \big[{\cal B}_i^n {\cal D}_n,{\cal
B}_k^m\big] {\cal D}_m=
\sum\limits_{n,m=1}^N{\cal B}_i^n{\cal B}_k^m \big[{\cal D}_n,
{\cal D}_m\big]+
\nonumber\\
+\sum\limits_{n,m=1}^N{\cal B}_i^n \big[{\cal D}_n,{\cal
B}_k^m\big] {\cal D}_m=
\sum\limits_{n,m=1}^N{\cal B}_i^n \big[{\cal D}_n,{\cal B}_k^m 
{\cal D}_m\big]=  
\sum\limits_{n=1}^N{\cal B}_i^n \big[{\cal D}_n,{\cal A}_k\big]+ 
\nonumber\\
+\sum\limits_{n=1}^N \big[{\cal B}_i^n,{\cal B}_k^0\big]{\cal D}_n=
\sum\limits_{n=1}^N \big[{\cal B}_i^n {\cal D}_n,{\cal A}_k\big]
=\big[{\cal A}_i,{\cal A}_k\big]=0
\label{im.11}
\end{eqnarray}
Combining (\ref{im.9}) and (\ref{im.11}), we obtain finally
\begin{eqnarray}
\sum\limits_{n,m=1}^N{\cal B}_i^n{\cal B}_k^m
\big[{\cal D}_n,{\cal D}_m\big]=0,\quad
\mbox{for~any~}i~\mbox{and~}k.
\label{im.12}
\end{eqnarray}
Now note that the invertibility of the operator $\rho$
implies invertibility of the matrix operator (\ref{mps.5a}).
Using this fact, we obtain the relations
\begin{eqnarray}
\big[{\cal D}_n,{\cal D}_m\big]=0,\quad
\mbox{for~any~}n~\mbox{and~}m,
\label{im.13}
\end{eqnarray}
i.e. commutativity of the operators
${\cal D}_i$ in the strong sense.
Thus, we proved the following theorem:

\medskip
{\bf Theorem 2.2.} If the system of multi-parameter
spectral equations (\ref{mps.3}) is non-degenerate and
commutative then the operators ${\cal D}_i,\ i=1,\ldots,N$
defined by formulas (\ref{im.1}) form a commutative family
and can be considered as integrals of motion of a certain
completely integrable quantum system. These operators have
a common set of eigenfunctions $\varphi$ constructed from
the solutions of system (\ref{mps.3}) and their eigenvalues
coincide with the values of the spectral parameters
$\lambda_i,\ i=1,\ldots,N$ of equation (\ref{mps.3}).

\subsection{Transformation properties of completely
integrable equations}

It is quite obvious that equivalence transformations do
not affect the invertibility properties of the operator $\rho$,
as follows from formula (\ref{mps.12d}).
However, these transformations may change the commutativity 
properties of the operators 
$B_i^j$. In other words, the commutativity of a certain system of
multi-parameter spectral equations does not automatically
imply the commutativity of its transformed version.
However, it turns out that the transformed version of
equations (\ref{im.2}) describes again a certain completely
integrable system. Indeed, applying to (\ref{im.2}) some equivalence
transformation, we obtain
\begin{eqnarray}
\tilde{\cal D}_j\tilde\varphi=\tilde\lambda_j\tilde\varphi, 
\qquad \tilde\varphi\in {\cal
V},\qquad j=1,\ldots,N
\label{im.13a}
\end{eqnarray}
where the transformed operators $\tilde{\cal D}_j$, using
formulas (\ref{im.1}), (\ref{mps.12c}) and (\ref{mps.12d}),
read as
\begin{eqnarray}
\tilde{\cal D}_j=
{\cal R}\left[\sum_{k=1}^N M_j^k({\cal D}_k-\mu_k I) \right]{\cal R}^{-1}
\label{im.13b}
\end{eqnarray}
But from (\ref{im.13}) and
(\ref{im.13b}) it immediately follows that for any $n$ and $m$
\begin{eqnarray}
\big[\tilde{\cal D}_n,\tilde{\cal D}_m\big]=0,
\label{im.13c}
\end{eqnarray}
i.e. the transformed model is again integrable.
This leads us to an important conclusion which
we formulate in the form of the following theorem:

\medskip
{\bf Theorem 2.3.} Let a system of multi-parameter spectral
equations be non-degenerate and equivalent to a certain 
commutative system. Then the operators ${\cal D}_i,\ i=1,\ldots,N$
defined by formulas (\ref{im.1}) form a commutative family
and can be considered as integrals of motion of a certain
completely integrable quantum system.

\subsection{Hermitian completely integrable systems}

Let us now consider a system of hermitian multi-parameter
spectral equations. Remember that hermiticity means
that the spaces $V_i$ are endowed with scalar 
products $(\ ,\ )_i$ with respect to which the operators
$A_i$ and $B_i^j$ are hermitian in $V_i$. Then formulas
(\ref{mps.4}) allow one
to introduce in a natural way a scalar product $\langle \ ,\ \rangle$ 
in the space ${\cal V}$. The operators ${\cal B}_i^j$ will obviously be 
hermitian in ${\cal V}$. Using the fact that the operator ${\cal B}_i$ 
commutes with all the operators ${\cal B}_k^j$ with $k\neq i$ and 
observing the absence of ${\cal B}_i^n$ in expression (\ref{mps.10}), 
we can conclude
\begin{eqnarray}
\left[\Delta{\cal B}_j^i,{\cal B}_i\right]=0,
\quad\mbox{for~any~}j\mbox{~and~}i.
\label{mps.14}
\end{eqnarray}
Note also that, due to the commutativity of the 
operators ${\cal B}_i^j$ and
their hermiticity in $V_i$, the operators 
(\ref{mps.8}) and (\ref{mps.10}) are
hermitian in ${\cal V}$:
\begin{eqnarray}
(\det {\cal B})^+=\det {\cal B},\qquad
(\bar{\cal B}_j^i)^+=\bar{\cal B}_j^i.
\label{mps.15}
\end{eqnarray}
From (\ref{mps.14}) and (\ref{mps.15}) it follows that the operators
${\cal C}_j$ and $\rho$ are hermitian in the space ${\cal V}$.

\medskip
{\bf Definition 2.4.} We call a system of multi-parameter
spectral equations {\it positive definite} if it has a
positive definite weight operators.

\medskip
It is obvious that positive definiteness of a hermitian
operator $\rho$ implies its invertibility and also the existence of
the square root $\rho^{1/2}$ which can also be considered an hermitian
operator. Let $\rho$ be a positive definite hermitian
operator. Then, after introducing new vectors $\Phi$ as 
\begin{eqnarray}
\Phi=\rho^{1/2}\varphi,
\label{pko.1}
\end{eqnarray}
and new operators $\hat{\cal D}_j$ as 
\begin{eqnarray}
\hat{\cal D}_j=
\rho^{-1/2}{\cal C}_j\rho^{-1/2},
\label{pko.2}
\end{eqnarray} 
equation (\ref{mps.11}) takes the form
\begin{eqnarray}
\hat{\cal D}_j\Phi=\lambda_j\Phi, \qquad \Phi\in {\cal V},\qquad j=1,\ldots,N
\label{mps.16}
\end{eqnarray}
The operators $\hat{\cal D}_j$ are obviously hermitian in the
space ${\cal V}$ and their eigenvalues in ${\cal V}$
coincide with the admissible values of the spectral parameters $\lambda_j$.
It is not difficult to see that the operators
(\ref{pko.2}) are related to the operators introduced in (\ref{im.1}) by
the formula
\begin{eqnarray}
\hat{\cal D}_j=\rho^{1/2}{\cal D}_j\rho^{-1/2}.
\label{mps.16aa}
\end{eqnarray}
From this formula it follows that commutativity of the
operators ${\cal D}_j$ implies commutativity of the
operators $\hat{\cal D}$. But from previous results 
we know that commutativity of the
operators ${\cal D}$ is implied by commutativity of the
initial system of multi-parameter spectral equations.
Collecting all these facts and recalling the definition of 
exact and quasi-exact solvability we arrive at the following 
important theorem:

\medskip
{\bf Theorem 2.4.} Let a system of multi-parameter spectral
equations be hermitian, positive definite and commutative.
Then the operators $\hat{\cal D}_j,\
j=1,\ldots,N$ are hermitian, form a commutative family, $[\hat{\cal
D}_j,\hat{\cal D}_k]=0$, and thus, can be considered as
``hamiltonians'' of a certain comletely integrable quantum system. 
If, in addition, the system of multi-parameter spectral equations is exactly
or quasi-exactly solvable, then the resulting completely
integrable system also will be exactly or quasi-exactly solvable.
 
\medskip
This theorem will play a central role in our further considerations.
Before completing this section, it is important to discuss
the case when the
initial multi-parameter spectral equation becomes degenerate.

\subsection{The degenerate case}

Assume that the operator $\rho$ is non-negative
definite and hence singular in ${\cal V}$. 
In this case, for any $j=1,\ldots,N$ 
there exists an orthogonal decomposition of the space
${\cal V}$ of the form
\begin{eqnarray}
{\cal V}={\cal V}^1\oplus{\cal V}^{2}_j\oplus{\cal V}^{3}_j
\label{mps.16a}
\end{eqnarray}
in which the operators $\hat{\cal C}_j$ and $\rho$ take the
following block form
\begin{eqnarray}
{\cal C}_j=
\left(
\begin{array}{ccc}
{\cal C}_j^{11} & {\cal C}_j^{12} & {\cal C}_j^{13} \\[0.2cm]
{\cal C}_j^{21} & {\cal C}_j^{22} & 0   \\[0.2cm]
{\cal C}_j^{31} & 0   & 0            
\end{array}
\right), \qquad
\rho =
\left(
\begin{array}{ccc}
r & 0 & 0\\[0.2cm]
0         & 0 & 0\\[0.2cm]
0         & 0 & 0
\end{array}
\right).
\label{mps.17}
\end{eqnarray}
Here $r \in \mbox{End}[{\cal V}^1]$ is a positive definite 
hermitian operator, ${\cal C}_j^{11} \in \mbox{End}[{\cal V}^1]$ and
${\cal C}_j^{22}\in \mbox{End}[{\cal V}^{2}_j]$ are hermitian 
operators, and ${\cal C}_j^{21} \in \mbox{Hom}[{\cal V}^{2}_j,{\cal V}^{1}]$
and ${\cal C}_j^{31}\in \mbox{Hom}[{\cal V}^{3}_j,{\cal V}^1]$ 
are hermitian conjugates of the operators 
${\cal C}_j^{12}\in \mbox{Hom}[{\cal V}^1,{\cal V}^{2}_j]$ and 
${\cal C}_j^{13}\in \mbox{Hom}[{\cal V}^1,{\cal V}^{3}_j]$. 
The decomposition (\ref{mps.16a}) can always be chosen in such a way that 
all the operators ${\cal C}_j^{22}$ are invertible. Writing
\begin{eqnarray}
\varphi = \left(
\begin{array}{c}
\varphi^1\\[0.2cm]
\varphi_j^2\\[0.2cm]
\varphi_j^3
\end{array}
\right)
\label{mps.18}
\end{eqnarray}
with $\varphi^1\in {\cal V}^1$, $\varphi_j^2\in {\cal V}^{2}_j$ and 
$\varphi_j^3\in {\cal V}^{3}_j$,
we can rewrite equations (\ref{mps.11}) in the form
\begin{eqnarray}
\left.
\begin{array}{rcl}
{\cal C}^{11}_j\varphi^1+{\cal C}^{12}_j\varphi^2_j+ {\cal
C}^{13}_j \varphi_j^{3}
&=& \lambda_j r\varphi^1 \\[0.2cm]
{\cal C}^{21}_j\varphi^1+{\cal C}^{22}_j\varphi^2_j  &=&  0 \\[0.2cm]
{\cal C}^{31}_j\varphi^1 & =&  0
\end{array}
\right\},
\qquad j=1,\ldots, N.
\label{mps.19}
\end{eqnarray}
From the last equation in (\ref{mps.19}) it follows that
\begin{eqnarray}
\varphi^1=P\varphi'
\label{mps.20}
\end{eqnarray}
where $P$ is a certain operator such that 
\begin{eqnarray}
{\cal C}^{31}_j P=0, \quad   P^+ {\cal C}^{13}_j=0,\quad
j=1,\ldots, N
\label{mps.21}
\end{eqnarray}
and $\varphi'\in {\cal V}'$ is a vector belonging to a certain
subspace ${\cal V}'\subset{\cal V}^1$. 
Substituting (\ref{mps.20}) into the
second equation in (\ref{mps.19}) and using (\ref{mps.21})
we obtain
\begin{eqnarray}
\varphi^2_j=-({\cal C}^{22}_j)^{-1}{\cal C}^{21}_j P\varphi'
\label{mps.22}
\end{eqnarray}
Substituting (\ref{mps.20}) and (\ref{mps.22}) into the
first equation in (\ref{mps.19}) and acting on the resulting
relation by the operator $P^+$ we obtain the following set
of one-parameter spectral equations
\begin{eqnarray}
{\cal C}'_j\varphi'=\lambda_j \rho'\varphi',\quad
\varphi'\in {\cal V}'
\label{mps.23}
\end{eqnarray}
in which
\begin{eqnarray}
{\cal C}'_j=P^+\left[{\cal C}^{11}_j-{\cal C}^{12}_j 
({\cal C}^{22}_j)^{-1} {\cal C}^{21}_j \right]P
\label{mps.24}
\end{eqnarray}
are hermitian operators in ${\cal V}'$ and
\begin{eqnarray}
\rho'=P^+ R P 
\label{mps.25}
\end{eqnarray}
is hermitian and positive definite in ${\cal V}'$.
Noting that $(\rho')^{1/2}$ is an
invertible hermitian operator and 
introducing the vectors, 
\begin{eqnarray}
\Phi'=(\rho')^{1/2}\varphi',
\label{mps.25a}
\end{eqnarray}
and the operators 
\begin{eqnarray}
\hat{\cal D}'_j = (\rho')^{-1/2}{\cal C}'_j(\rho')^{-1/2},
\label{mps.25b}
\end{eqnarray} 
we can reduce equations (\ref{mps.23}) to the form
\begin{eqnarray}
\hat{\cal D}'_j\Phi'=\lambda_j\Phi', \qquad 
\Phi'\in {\cal V}',\qquad j=1,\ldots,N
\label{mps.26}
\end{eqnarray}
The operators $\hat{\cal D}'_j$ are obviously hermitian in the
space ${\cal V}'$ and their eigenvalues in ${\cal V}'$
coincide with the admissible values of the spectral parameters $\lambda_j$.
As before, if the set of solutions $\Phi'$ is complete in ${\cal V}'$,
the operators $\hat{\cal D}'_j,\
j=1,\ldots,N$ form a commutative family, $[\hat{\cal
D}'_j,\hat{\cal D}'_k]=0$,
and thus, can be considered as ``hamiltonians'' of a
certain completely integrable quantum system. Unfortunately,
at present we do not have more general criteria for
commutativity of the operators ${\cal D}'_j$ in the degenerate case.

\section{Conclusion. Exact and quasi-exact solvability}

Let us now discuss some solvability properties of systems of
multi-parameter spectral equations. First note that if all
the carrier spaces $V_i$ are finite-dimensional, then the
solution of the system of multi-parameter spectral
equations is a purely algebraic problem (see e.g. examples
1.1 and 1.3). In  the case of infinite-dimensional carrier
spaces an algebraic solution of the system of 
multi-parameter spectral equations is,
as a rule, impossible (see e.g. example 1.4). However, 
the presence of some high symmetry
in the system may turn it algebraically solvable,
despite the infinite-dimensionality of the carrier spaces.

\medskip
{\bf Definition 4.1.} We call an infinite-dimensional 
system of multi-parameter spectral
equations {\it exactly solvable} if all its solutions can be
constructed algebraically. We call an infinite-dimensional 
system of multi-parameter
spectral equations {\it quasi-exactly solvable} if only a
certain finite part of its solutions allows an algebraic construction.

\medskip 
Consider an infinite-dimensional system of 
$N$-parameter spectral equations (\ref{mps.3}) which is
known to be exactly solvable. Assume that the spectrum of
this system is partially degenerate and thus,
by definition, contains at least one 
finite, degenerate and non-extendable subset $\Sigma'$.
Denoting its degree of degeneracy by $K$, one can write
\begin{eqnarray}
\left[A_i-\sum_{j=1}^N B_i^j \lambda_j\right]
\phi_i=0, 
\quad V_i\ni \phi_i \neq 0,\quad
(\lambda_1,\ldots,\lambda_N) \in \Sigma',  \quad i=1,\ldots,N.
\label{mps.3p}
\end{eqnarray}
According to the results of the previous subsection, 
the degeneracy allows to apply 
to relations (\ref{mps.3p}) an equivalence transformation,
which turns them into the form
\begin{eqnarray}
\left[\tilde A_i-\sum_{j=1}^{N'}\tilde B_i^j \tilde\lambda_j
\right]
\tilde\phi_i=0, 
\quad V_i\ni \tilde\phi_i \neq 0,\quad 
(\lambda_1,\ldots,\lambda_{N'}) \in \tilde\Sigma',
\quad i=1,\ldots,N',
\label{mps.3q}
\end{eqnarray}
Here we used $\tilde\lambda_j=0$ for $j=N-K+1,\ldots,N$,
suppressed the last $K$ equations for $i=N-K+1,\ldots,N$
and introduced the notation $N'=N-K$. We also denoted by $\tilde\Sigma'$
the set of the transformed spectral points with the zero components
suppressed.
It is easily seen that system (\ref{mps.3q}) is nothing else but the
system of $N'$-parameter spectral equations whose spectrum contains
by construction a certain finite exactly solvable part $\tilde\Sigma'$. 
Since the tranformed system (\ref{mps.3q}) is infinite-dimensional, its
total spectrum $\tilde\Sigma$ is infinite. 
This means that the $N'$-parameter spectral equation 
obtained this way is quasi-exactly solvable. Thus we arrive at the following
theorem:

\medskip
{\bf Theorem 4.2.} Any infinite-dimensional and exactly
solvable system of $N$-parameter
spectral equations with partially degenerate spectrum 
can be reduced to a certain quasi-exactly solvable $N'$-parameter
spectral equation. Here $N'=N-K$ where $K$ denotes the 
degree of the degeneracy of the spectrum of
the initial equations.

\end{document}